\documentstyle[prc,aps,preprint,psfig]{revtex}
\begin{document}
\title{$e^+e^-$ pair production from $\gamma $A reactions}
\author{M. Effenberger\footnote{part of Ph.D. thesis of M. Effenberger},
E. L. Bratkovskaya and U. Mosel \\[5mm]
Institut f\"ur Theoretische Physik,
Universit\"at Giessen\\D-35392 Giessen, Germany }
\maketitle

\begin{abstract}
We present a calculation of $e^+e^-$ production in $\gamma A$
reactions at MAMI and TJNAF energies within a semi-classical BUU transport 
model.
Dilepton invariant mass spectra 
for $\gamma$C, $\gamma$Ca and $\gamma$Pb are calculated at
0.8, 1.5 and 2.2 GeV.
We focus on observable effects of medium modifications of the $\rho$
and $\omega$ mesons. The in-medium widths of these mesons are taken into
account in a dynamical, consistent way. We discuss the transport 
theoretical treatment of broad resonances.   
\end{abstract}

\vspace*{1cm}


\newpage
\section{Introduction}

In-medium properties of hadrons are of fundamental interest with respect
to an understanding of QCD in the non-perturbative regime (cf. \cite{kochko}).
During the past decade especially the properties of vector mesons have
found widespread attention as they may be related to chiral symmetry 
\cite{mosel91,BrownRho,H&L92}.     
\par A comparison of experimental data on dilepton production in 
nucleus-nucleus collisions at SPS energies \cite{CERES,HELIOS} with
transport theoretical calculations \cite{likobrown,CBRep98} seems to indicate 
a lowering of the $\rho$-meson mass in the nuclear medium.
However, since in a heavy ion collision the final dilepton yield is
obtained by an integration over dileptons emitted at different densities and
temperatures a discrimination between different scenarios of in-medium
modifications for the vector mesons is difficult \cite{kochsong,CBRW97}. 
\par Moreover,
there is a more fundamental concern: ultrarelativistic heavy ion reactions
proceed, at least in their initial stages, far from equilibrium, whereas all
theoretical predictions of in-medium properties are based on equilibrium
assumptions.
Therefore it is necessary to probe in-medium properties of vector mesons
under 'cleaner' conditions. For that purpose photon or pion induced reactions
are promising tools. In such reactions the nuclear
medium is very close to equilibrium at normal nuclear density and temperature 
zero. The predicted in-medium effects for the vector mesons are so large that
they should have observable consequences already for densities $\rho=\rho_0$.
Here one should note that even the dileptons seen in ultrarelativistic heavy ion
reactions stem to a large part from densities $\rho \le 2 \rho_0$.
\par Our calculations are based on a semi-classical BUU transport model that
has recently been very successfully applied to the description of heavy ion 
collisions at SIS energies \cite{TeisZP97} and photoproduction of pions and
etas in nuclei \cite{prodpaper}. Meanwhile we have extended the model
to the description of dilepton as well as strangeness production and to the
high energy regime by including the Lund string fragmentation model FRITIOF 
\cite{FRITIOF}. This allows us to calculate inclusive particle production
in heavy ion collisions from 200 AMeV to 200 AGeV, in photon and in pion
induced reactions with the very same physical input. To our opinion, the 
simultaneous 
description of as many experimental observables as possible is necessary
because of the large number of parameters, like unknown cross sections, and
the strong assumptions that enter semi-classical transport models.
In heavy-ion collisions particle production depends not only on a large
number of elementary reaction channels but also on the global space-time
evolution of the system. On the other hand photon, pion or proton induced reactions allow to fix certain
ingredients of the transport model as observables depend in general only on
a few reaction channels. In Ref.~\cite{etan} we have, for example, determined
the in-medium $\eta$-nucleon cross section from photoproduction of 
$\eta$-mesons. Just recently we were able to improve considerably our 
treatment of the $\Delta$-resonance by comparing our calculations to
experimental data on photoproduction of pions \cite{piondel}. 
\par Within our model we have already given predictions for
dilepton production in pion nucleus reactions \cite{pidil} that will be 
measured by the HADES collaboration at GSI \cite{HADES}. In the present paper
we want to look at $e^+e^-$
production in photonuclear reactions in the energy range from 0.8 to 2.2 GeV
that will be accessible at TJNAF \cite{CEBAF} and at its lowest energy also 
at MAMI.
\par Our paper is organized as follows: In Section II we describe our model.
Here we focus on the treatment of broad resonances and our description 
of elementary photon nucleon reactions. In Section III we present our results
for dilepton production in photon nucleus reactions and discuss the 
possibility to subtract the Bethe-Heitler contribution.
In Section IV we include in-medium modifications for the vector mesons into
our calculations and present their effect on the dilepton yield. 
We close with a summary in Section V.

\section{The BUU model}

The transport model used here has been developed starting from the model that has been described in full detail in Refs.~\cite{TeisZP97,Effen97}. Here we restrict ourselves to the description of the essential new features of our method.

\subsection{Resonance properties}

Instead of the baryonic resonances in Ref.~\cite{TeisZP97} that were taken with their parameters from the PDG~\cite{PDG2} we now use the resonances from the analysis of Manley and Saleski~\cite{Manley}. This has the advantage of supplying us with a consistent set of resonances. In particular, now the resonance parameters are consistent with the parameterizations of the resonance widths. We take into account all resonances that are rated with at least 2 stars in Ref.~\cite{Manley}: $P_{33}$(1232), $P_{11}$(1440), $D_{13}$(1520), $S_{11}$(1535), $P_{33}$(1600), $S_{31}$(1620), $S_{11}$(1650), $D_{15}$(1675), $F_{15}$(1680), $P_{13}$(1879), $S_{31}$(1900), $F_{35}$(1905), $P_{31}$(1910), $D_{35}$(1930), $F_{37}$(1950), $F_{17}$(1990), $G_{17}$(2190), $D_{35}$(2350). The resonances couple to the following channels: $N \pi$, $N \eta$, $N \omega$, $\Lambda K$, $\Delta(1232) \pi$, $N \rho$, $N \sigma$, $N(1440) \pi$, $\Delta(1232) \rho$. The cross section for the production of a resonance $R$ in a collision of a meson $m$ with a baryon $B$ is given by:
\begin{equation}
\sigma_{m B \to R}=\frac{2 J_R+1}{(2 J_m + 1)(2 J_B +1)} \frac{4 \pi}{k^2} 
\frac{s \Gamma^{in}_{m B} \Gamma^{out}_{tot}}{(s-M_R^2)^2+s {\Gamma^{out}_{tot}}^2} \quad,
\label{ressig}
\end{equation}
where $J_R$, $J_m$, $J_B$ denote the spins of the resonance, the baryon and the meson, respectively. $k$ is the cms momentum of the incoming particles, $s$ is the squared invariant energy, and $M_R$ is the pole mass of the resonance. The total decay width $\Gamma^{out}_{tot}$ is given as sum over the partial decay widths of the resonance. For a specific channel $m B$ it is:
\begin{equation}
\Gamma^{out}_{m B}=\Gamma^0_{m B} \frac{\rho_{m B}(s)}{\rho_{m B}(M_R)} \quad,
\label{gamout}
\end{equation}
with $\Gamma^0_{m B}$ being the partial decay width at the pole of the resonance and $\rho_{m B}(s)$ is given as:
\begin{equation}
\rho_{m B}(s)=\int d\mu_m d\mu_B {\cal A}_m(\mu_m) {\cal A}_B(\mu_B) 
\frac{q(s,\mu_m,\mu_B)}{\sqrt{s}} B^2_{l_{m B}}(qR) \quad,
\label{gamps}
\end{equation}
where ${\cal A}_m$ and ${\cal A}_B$ denote the spectral functions of the outgoing particles. $q$ is their cms momentum, $l_{m B}$ is their relative orbital angular momentum, and $B_{l_{m B}}$ is a Blatt-Weisskopf barrier penetration factor \cite{Blatt,Manley} (interaction radius $R=1\, {\rm fm}$). For the spectral function ${\cal A}_i$ of an unstable particle $i$ we use:
\begin{equation}
{\cal A}_i(\mu)=\frac{2}{\pi} \frac{\mu^2 \Gamma_{tot}(\mu)}{(\mu^2-M_i^2)^2+\mu^2 \Gamma^2_{tot}(\mu)} \quad,
\label{spectral}
\end{equation}          
where $M_i$ denotes the pole mass and $\Gamma_{tot}(\mu)$ is the total width.
Here we neglect any spin degrees of freedom as well as a momentum dependence of the real part of the self-energy. 
For a stable  particle (with respect to the strong interaction) we simply have:
\begin{equation}
{\cal A}_i(\mu)=\delta(\mu-M_i) \quad.
\end{equation}
The incoming width $\Gamma^{in}_{m B}$ in Eq.~(\ref{ressig}) is given by:
\begin{equation}
\Gamma^{in}_{m B}=C^{I_R}_{m B} \Gamma^0_{m B} \frac{k B^2_{l_{m B}}(k R)}
{\sqrt{s} \rho_{m B}(M_R)} \quad,
\end{equation}
where $C^{I_R}_{m B}$ is the appropriate Clebsch-Gordan coefficient for the
coupling of the isospins of the baryon and the meson to the isospin $I_R$ of
the resonance.
\par We note that we use -- in contrast to Ref.~\cite{Manley} -- relativistic propagators in Eqs.~(\ref{ressig}),(\ref{spectral}) and momentum dependent widths in the spectral functions of the outgoing particles. However, this has only a very small effect on the resonance production cross sections and does not require a readjustment of the resonance parameters.   
\par The mesonic resonances are treated analogously to the baryonic ones, i.e. their two-body decay widths are calculated according to Eq.~(\ref{gamout}). The used parameters are given in Table~\ref{mestable}. The three pion decay width of the $\omega$-meson is assumed to be constant since it is very small.

\subsection{The collision term}

\subsubsection{Baryon-baryon collisions}
For invariant energies $\sqrt{s}<2.6 \, {\rm GeV}$ we describe baryon-baryon collisions as in Ref.~\cite{TeisZP97} with the same matrix elements. Our modified treatment of the resonance properties preserves the very good agreement with the experimental data on one- and two-pion production in nucleon-nucleon collisions shown in Ref.~\cite{TeisZP97}. For higher energies we use the string fragmentation model FRITIOF~\cite{FRITIOF} with the same parameters as in Ref.~\cite{geiss}. This approach is similar to the ones in the hadronic transport models
described in Refs.~\cite{hsd,urqmd}.

\subsubsection{Meson-baryon collisions}
For meson-baryon collisions we use the string fragmentation model for $\sqrt{s}>2.2 \, {\rm GeV}$. For lower energies the most important contributions come from intermediate nucleon resonances which are described according to Eq.~(\ref{ressig}). 
\par
\begin{figure}[h]
\centerline{\psfig{figure=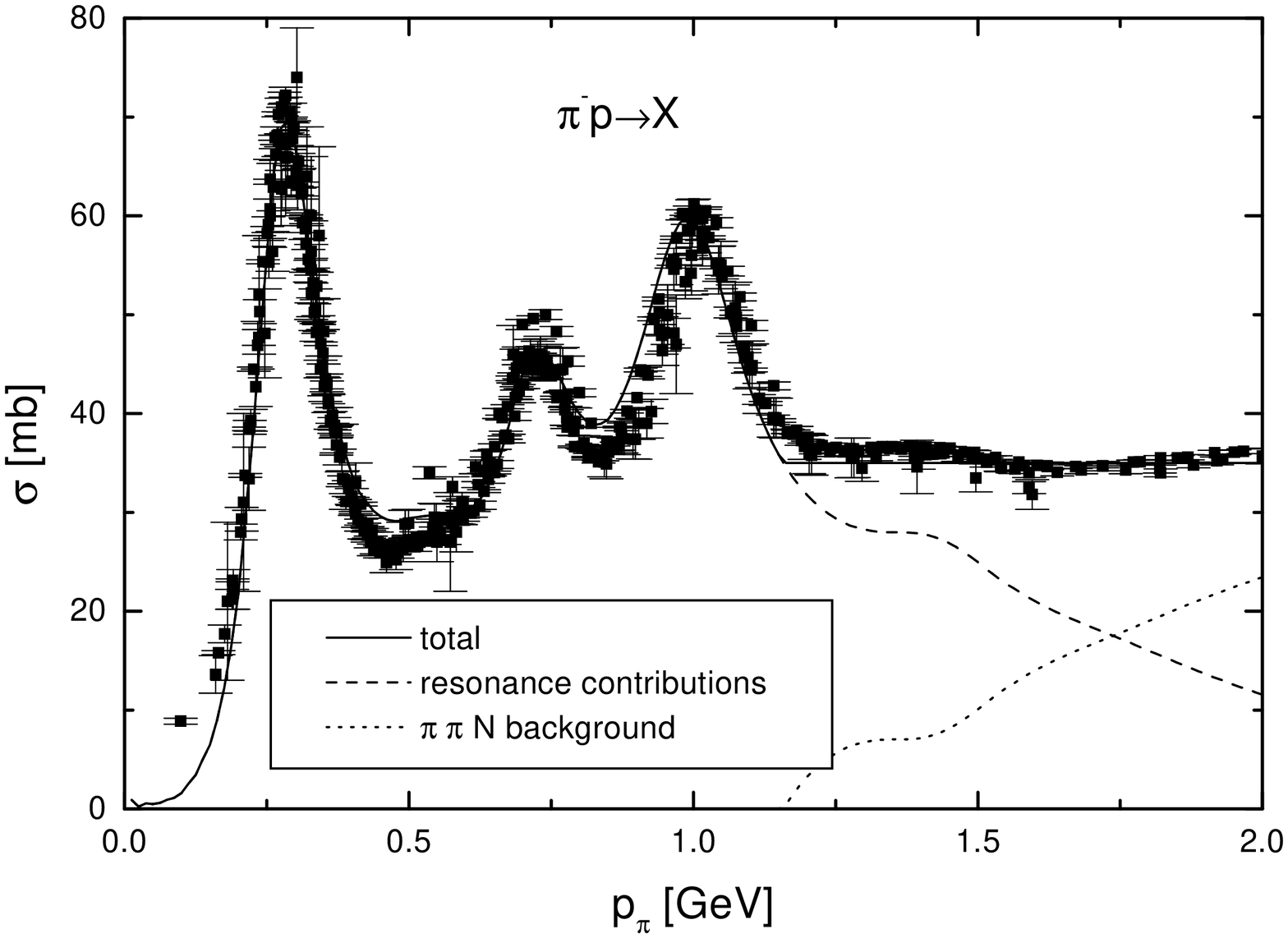,width=10cm}}
\caption{The total $\pi^-p$ cross section compared to the experimental data 
from \protect\cite{landolt}.}
\label{fig_pim_tot}
\end{figure}

\begin{figure}[h]
\centerline{\psfig{figure=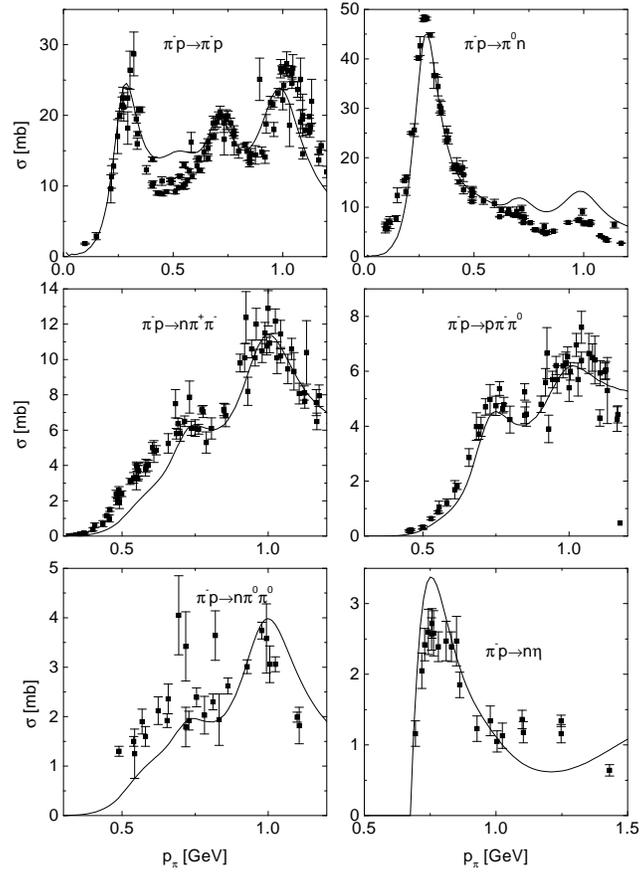,width=10cm}}
\caption{$\pi^-p$ cross sections for elastic scattering, charge exchange, two-pion and eta production. The experimental data are taken 
from \protect\cite{landolt}.}
\label{fig_pim_ex}
\end{figure}

 In case of $\pi^-p$-scattering the incoherent sum of all resonance contributions gives a very good agreement with experimental data for pion momenta $p_\pi \le 1.1 \, {\rm GeV}$, corresponding to invariant energies $\sqrt{s} \le 1.73 \, {\rm GeV}$. In Fig.~\ref{fig_pim_tot} we show the total $\pi^- p$ cross section.  
For higher energies ($1.73 \, {\rm GeV} < \sqrt{s} < 2.2 \, {\rm GeV}$)
we additionally include a background $\pi N \to \pi \pi N$ cross section in order to reproduce the total cross section for which we use a parameterization from the PDG~\cite{PDG94} (shown as solid line in Fig.~\ref{fig_pim_tot}).
In Fig.~\ref{fig_pim_ex} we show that the resonance contributions give a very good description of experimental data on elastic scattering, charge exchange, two-pion and eta production cross sections.
\par
\begin{figure}[h]
\centerline{\psfig{figure=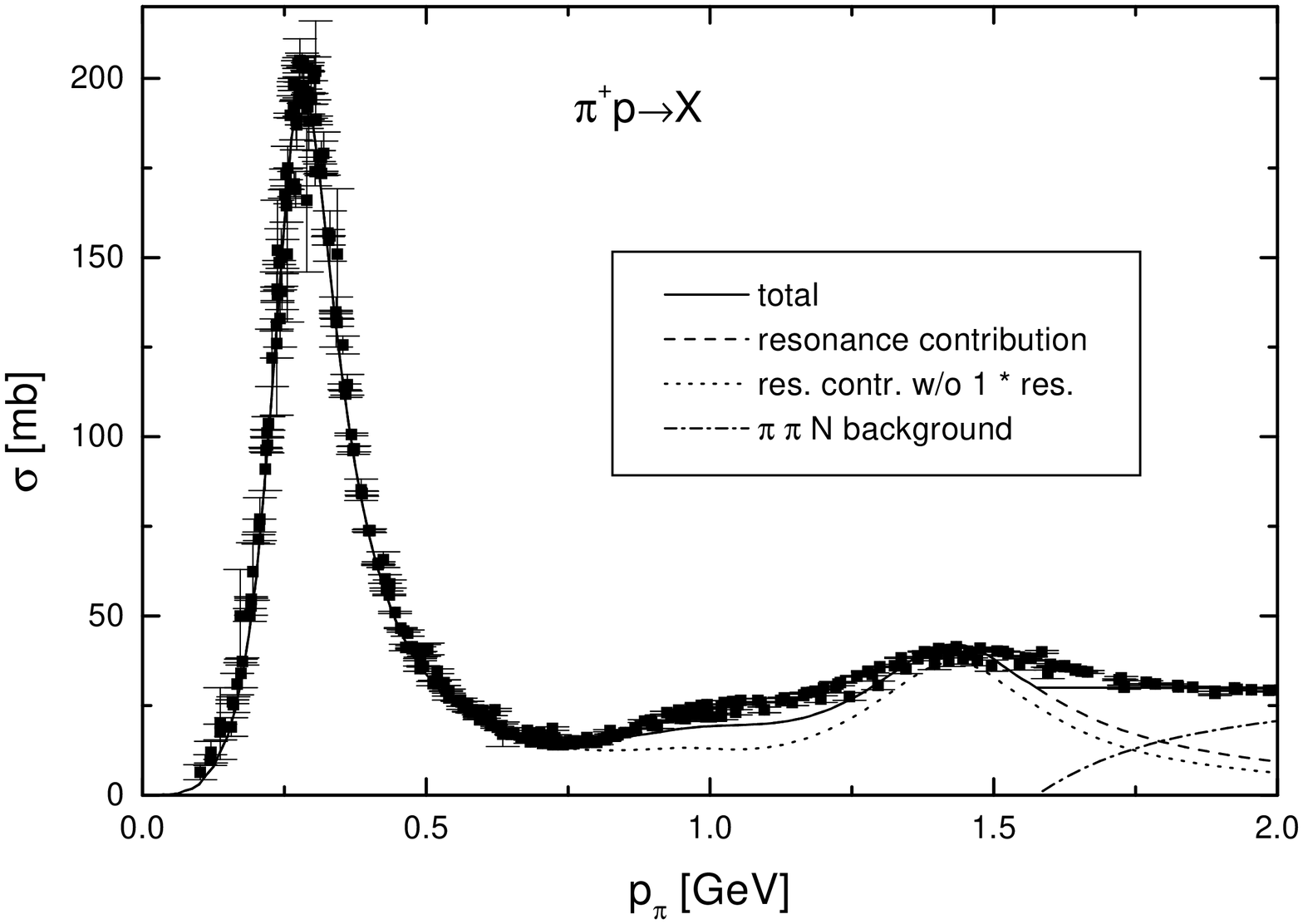,width=10cm}}
\caption{The total $\pi^+p$ cross section compared to the experimental data 
from \protect\cite{landolt}.}
\label{fig_pip_tot}
\end{figure}

\begin{figure}[h]
\centerline{\psfig{figure=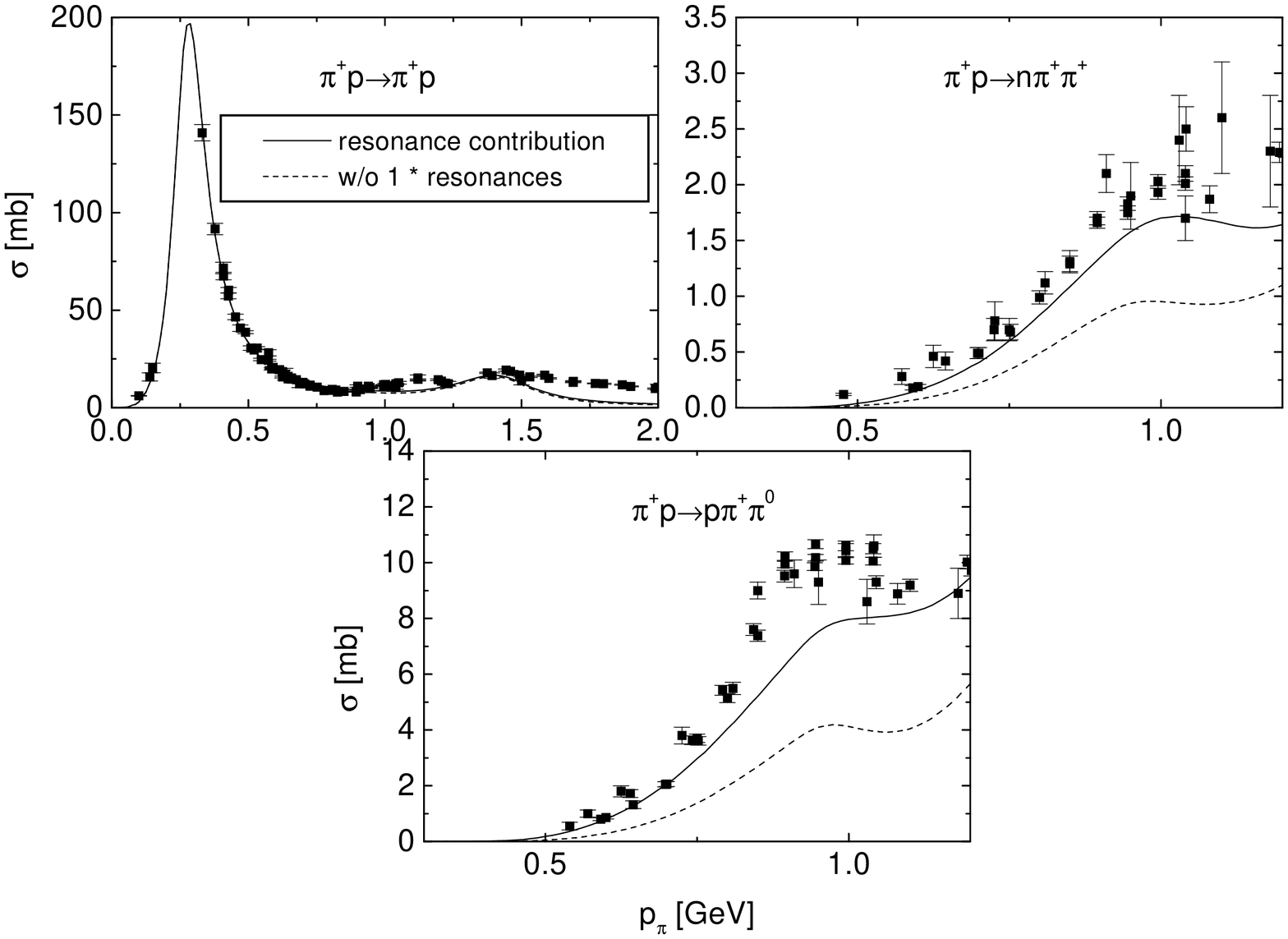,width=10cm}}
\caption{$\pi^+p$ cross sections for elastic scattering and two-pion production. The experimental data are taken from \protect\cite{landolt}.}
\label{fig_pip_ex}
\end{figure}

 For $\pi^+p$-scattering one gets, also for lower energies, a good agreement with experimental data only if one takes into account all resonances from Ref.~\cite{Manley}, including the 1 star ones, as can be seen from 
Figs.~\ref{fig_pip_tot} and \ref{fig_pip_ex}. Since we do not explicitly propagate the 1 star resonances we put their contributions into background cross sections for $\pi N \to \pi N$ and $\pi N \to \pi \pi N$. For higher energies we, again, include a two-pion production background term that is fitted to the total cross section from Ref.~\cite{PDG94}. 
\par For $\pi^0p$-scattering we have from isospin symmetry:
\[
\sigma_{\pi^0p}=\frac{1}{2}(\sigma_{\pi^+p}+\sigma_{\pi^-p})
\]
for the total cross section. The cross sections for pion-neutron scattering also follow from isospin symmetry.
\par
\begin{figure}[h]
\centerline{\psfig{figure=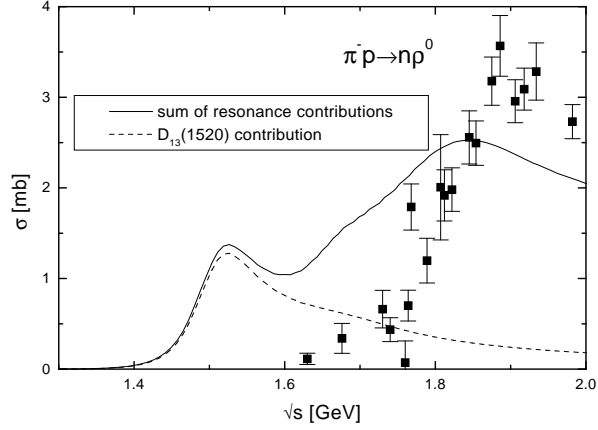,width=10cm}}
\caption{Cross section for $\pi^- p \to n \rho^0$. The experimental data are
taken from Ref.~\protect\cite{brody}.}
\label{pi_rho}
\end{figure}

 In Fig.~\ref{pi_rho} we compare the resonance contributions to 
$\pi^- p \to n \rho^0$ to the experimental data from Ref.~\cite{brody}. While
for invariant energies $\sqrt{s}$ above 1.8 GeV the experimental data are
reasonably well described by the resonances there is a strong disagreement at
lower energies, in particular in the region of the $D_{13}(1520)$-resonance.
The $\rho$-mesons produced at these energies have invariant masses essentially
below the pole mass $m_\rho^0$. In Ref.~\cite{brody} the $\rho$-meson
cross section has been obtained by a fit to invariant mass spectra of the
outgoing pions in $\pi^- p \to n \pi^+ \pi^-$. For low $\sqrt{s}$ the shapes
of the $\rho$-meson and the 'background' contributions become very similar
and a determination of the $\rho$-meson cross section gets very difficult.   
\par In Ref.~\cite{Manley}
the couplings of the baryonic resonances to the $N\rho$ channel have been
determined by using amplitudes for $\pi N \to N \rho$ that were obtained
by a partial wave analysis of all available $\pi N \to N \pi \pi$ data in
Ref.~\cite{Manley2}. We also note that the large coupling of the 
$D_{13}(1520)$-resonance to the $N\rho$ channel found in Ref.~\cite{Manley}
is in line with other similar analyses \cite{longacre}. Therefore we consider the
experimental data in Ref.~\cite{brody} to be wrong for low $\sqrt{s}$.        
\par In addition to the resonance contributions we include the following processes:
\begin{eqnarray*}
\pi N &\leftrightarrow& \omega N \\
\pi N &\to& \omega \pi N \\
\omega N &\to& \pi \pi N \\
\omega N &\to& \omega N \\
\pi N &\leftrightarrow& \phi N \\
\pi N &\to& \phi \pi N \\
\phi N &\to& \pi \pi N \\
\phi N &\to& \phi N \quad,
\end{eqnarray*}
where we adopt the cross sections from Ref.~\cite{golub}. For cross sections involving an $\omega$-meson we, of course, subtract our resonance contributions from these cross sections.   

\subsection{Treatment of broad resonances}
\label{broadres}
In our model broad resonances, like the baryon resonances or the $\rho$-meson, are not just produced and propagated with their pole mass but according to their spectral function. The transport equation for a system of $N$ particle species reads:
\begin{equation} 
(\frac{\partial}{\partial t} + \frac{\partial H_i}{\partial \vec{p}} \frac{\partial}{\partial \vec{r}} - \frac{\partial H_i}{\partial \vec{r}} \frac{\partial}{\partial \vec{p}})F_i=G_i {\cal A}_i - L_i F_i \quad (i=1,\ldots,N) \quad,
\label{transport}
\end{equation}
where $F_i(\vec{r},\vec{p},\mu,t)$ denotes the one-particle spectral phase space density of particle species $i$ with $\vec{r}$ and $\vec{p}$ being the spatial and momentum coordinates of the particle. $\mu$ is the invariant mass of the particle and $H_i(\vec{r},\vec{p},\mu,F_1,\ldots,F_N)$ stands for the single particle mean field Hamilton function which, in our numerical realization~\cite{TeisZP97}, is given as:
\begin{equation}
H_i=\sqrt{(\mu+S_i)^2+\vec{p}^2} \quad,
\end{equation}
where $S_i(\vec{r}, \vec{p}, \mu, F_1, \ldots, F_N)$ is a scalar potential.
We note that we neglect a vector potential and $S_i$ is an effective scalar
potential that is -- for the nucleons -- obtained from a non-relativistic
potential (for details see Ref.~\cite{TeisZP97}). 
The terms $G_i(\vec{r}, \vec{p}, \mu, F_1, \ldots, F_N)$ and $L_i(\vec{r}, \vec{p}, \mu, F_1, \ldots, F_N)$ 
stand for a gain and a loss term, respectively, and ${\cal A}_i(\vec{r},\vec{p},\mu,F_1,\ldots,F_N)$ is the spectral function of particle $i$. The distribution function $f_i$ is defined by:
\begin{equation}
f_i(\vec{r},\vec{p},\mu,t)=\frac{F_i(\vec{r},\vec{p},\mu,t)}{{\cal A}_i(\vec{r},\vec{p},\mu,t)}  \quad;
\label{smallf}
\end{equation}
for stable particles it reduces to the usual phase space density. 
\par In order to be more specific about $G_i$ and $L_i$ let us consider, as an example, a system of nucleons, rho-mesons, pions and a single baryonic resonance species R that are coupled only via $R \leftrightarrow N \rho$ and $\rho \to \pi \pi$. Then the gain term $G_\rho$ is:
\begin{equation}
G_\rho = \frac{1}{{\cal A}_\rho} \int \frac{d^3 p_R}{(2 \pi)^3} d\mu_R F_R(\vec{r},\vec{p}_R,\mu_R,t) \frac{d \Gamma_{R \to N \rho}}{d^3 p_\rho d\mu_\rho} (1-f_n(\vec{r},\vec{p}_n,t)) \quad,
\label{gain}
\end{equation}
where the factor $(1-f_n)$ accounts for the Pauli principle of the outgoing nucleon and $\vec{p}_n=\vec{p}_R-\vec{p}_\rho$ due to momentum conservation. For simplicity, we have neglected a possible finite width of the nucleons as well as a Bose enhancement factor $(1+f_\rho)$ for the $\rho$-meson. In the differential decay width $d\Gamma_{R \to N \rho}$ the spectral function of the $\rho$-meson ${\cal A}_\rho$ enters simply as a multiplicative factor (Eqs.~(\ref{gamout}),(\ref{gamps})). Therefore, here $G_\rho$ does not depend on ${\cal A}_\rho$ since we neglect any process where more than one $\rho$-meson is produced, like e.g. $R \to \rho \rho N$.  
\par The loss term $L_\rho$ reads:
\begin{equation}
L_\rho=\Gamma_{\rho \to \pi \pi}+\int \frac{d^3p_n}{(2 \pi)^3} f_n(\vec{r},\vec{p}_n,t) v_{n \rho} \sigma_{\rho n \to R} \quad,
\label{loss}
\end{equation}
with $v_{n \rho}$ being the relative velocity of the nucleon and the $\rho$-meson and $\sigma_{\rho n \to R}$ their cross section for the production of resonance $R$ (Eq.~(\ref{ressig})). $\Gamma_{\rho \to \pi \pi}$ denotes the two-pion decay width of the $\rho$-meson in the calculational frame.
\par The total in-medium width $\Gamma_{tot,\rho}$ appearing in the spectral function from Eq.~(\ref{spectral}) is directly related to the loss rate $L_\rho$:
\begin{equation}
\Gamma_{tot,\rho}=\gamma L_\rho \quad,
\label{consist}
\end{equation}
where $\gamma$ is a Lorentz factor which appears since $\Gamma_{tot,\rho}$ is the decay rate in the rest frame of the $\rho$-meson. 
\par The loss and gain term for the resonance $R$ can be written down in an analogous way. One immediately sees that the transport equations of the $\rho$-meson and the resonance $R$ are coupled in a higly non-linear way. Especially the in-medium widths, that are functions of space time and 4-momentum, of both particles depend on each other through integral equations (Eqs.~(\ref{ressig}),(\ref{gamout}),(\ref{gamps}),(\ref{spectral}),(\ref{loss})).    
\par The above described equations can easily be extended to a transport model with more particle species and a realistic collision term. We also note that it is straightforward to formulate the theory consistently for real and imaginary part of the self-energies of the particles. However, in our model we treat real and imaginary part of the self-energies completely independently. This violates analyticity, but in the present stage it would already require a considerable effort to treat the imaginary parts of all particles in a realistic transport model in a completely self-consistent way.
\par In Ref.~\cite{knoll} it has recently been stressed that, because of 
unitarity, it is important to respect Eq.~(\ref{consist}) in transport 
calculations. This means that in the population of a resonance the same
width has to be used in the spectral function that enters the dynamical calculation via the collision term. 
In Refs.~\cite{prodpaper,ehehalt} we have already taken into account the
in-medium widths of the baryonic resonances $\Delta(1232)$, $N(1520)$, $N(1535)$, and $N(1680)$ in a consistent way for population and propagation. 
Since during a photoproduction reaction the nucleus remains, in the time interval relevant for meson production, close to its ground state the calculational effort is in this case manageable. As we describe the nuclear ground state in a local Thomas-Fermi approximation we can use nuclear matter values for the in-medium widths that depend only on the invariant mass $\mu$, the absolute value of the 3-momentum $|\vec{p}|$, and the density $\rho$:
\[
\Gamma(\vec{r},t,\vec{p},\mu) \to \Gamma(\rho(\vec{r},t),|\vec{p}|,\mu) \quad.
\] 
\par In the present paper we only take into account the in-medium widths of the $\rho$- and $\omega$-mesons. In particular, we neglect any medium-modifications of the $N \rho$-widths of the baryonic resonances that would, in a self-consistent calculation of the self-energies, directly follow from a modification of the spectral function of the $\rho$-meson. 
As was shown in Ref.~\cite{peters} such effects might give large modifications
of the baryon widths and also influence the spectral function of the 
$\rho$-meson. However, an inclusion of such effects would enhance the 
numerical effort dramatically, especially if one takes also the modifications
of the real parts of the self-energies into account. Moreover, we do not
expect such effects to be relevant for photon energies above 1 GeV because
here the nucleon resonances that lie below $M_N+m^0_\rho$ and might thus get
strongly modified, like the $D_{13}(1520)$, play only a minor role. 
Observable effects of different medium modifications of the $D_{13}$ will be
reported elsewhere.   
\par The transport equations Eq.~(\ref{transport}) do not yet give the correct asymptotic spectral phase space densities for particles that are stable in vacuum. 
This can be seen by noting that a collision broadened particle does not automatically lose its collisional width when being propagated out of the nuclear environment.
The same problem appears for resonances whose imaginary part of the self-energy is non-zero in-medium in kinematical regimes where it is zero in vacuum. 
\par The reason for this deficiency is directly related to the semi-classical approximations on which the transport equation Eq.~(\ref{transport}) is based, in particular the neglect of coherence effects. 
However, since a realistic quantum transport theory is numerically not yet realizable we will nevertheless work with Eq.~(\ref{transport}) as it is certainly a step beyond the usual on-shell approximation. Moreover, if the time evolution of the system is such that the rate of particle production and absorption
is much larger than the change of the width with time
the problem with surviving off-shell contributions will be negligible. Under
the assumption that the gain term in the collision term is in magnitude 
comparable to the loss term we can formulate this in the following way:  
\begin{equation}
\Gamma dt \gg \frac{d \Gamma}{\Gamma} \quad.
\end{equation}                    
In case of a particle moving in a static nuclear medium with density profile $\rho(\vec{r})$ we can rewrite this condition:
\begin{equation}
\frac{\vec{\nabla}\rho \cdot \vec{e}_p}{\rho} \ll \frac{1}{\lambda} \quad,
\label{condition}
\end{equation}
where $\vec{e}_p$ is a unit vector along the momentum direction of the particle and $\lambda$ denotes its mean free path.
In Section~\ref{collbroad} we will discuss the validity of this condition for our calculations and present possible remedies to the problem. 

\subsection{Parameterization of the elementary $\gamma N$ cross sections}
\label{gamele}
\begin{figure}[h]
\centerline{\psfig{figure=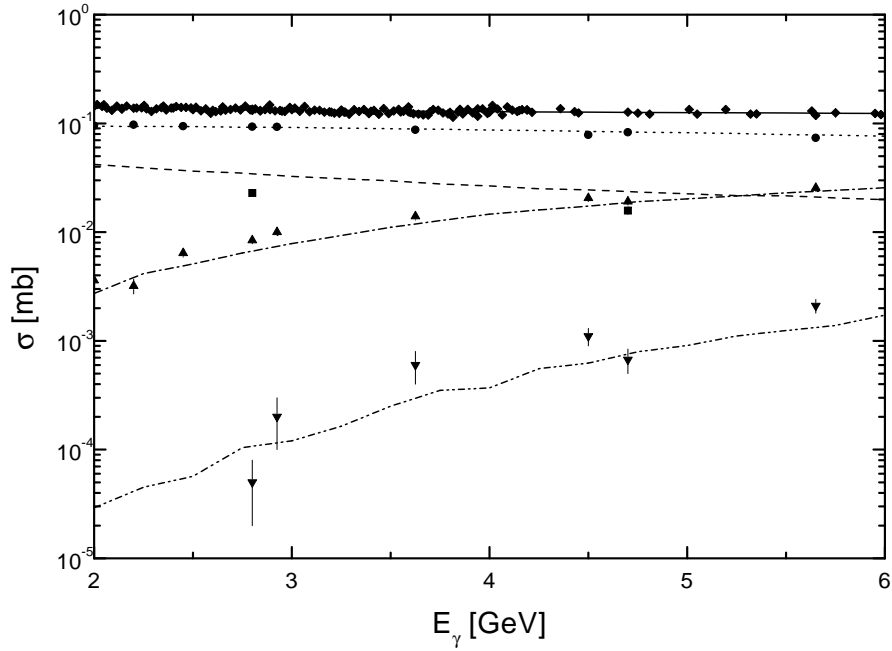,width=15cm}}
\caption{Charged particle multiplicity cross sections in $\gamma p$ reactions: 
$\gamma p \to$ 1 charged particle (dashed line (calculation as described in the text), squares (experimental data from Ref.~\protect\cite{landolt})), 3 charged particles (dotted line, circles), 5 charged particles (dot-dashed line, up triangles), 7 charged particles (dot-dot-dashed line, down triangles). Also shown is the total cross section (solid line, rhombs).}
\label{chargemult}
\end{figure}

For invariant energies $\sqrt{s}<2.1\,{\rm GeV}$, corresponding to $E_\gamma<1.88\,{\rm GeV}$ on a free nucleon at rest, we describe one-pion, two-pion and eta production as in Ref.~\cite{prodpaper}. For the two-pion production cross sections on the neutron we meanwhile use the experimental data from Refs.~\cite{TAPS,Daphne} instead of the recipe described in Ref.~\cite{prodpaper}. For larger energies we use, like for the hadronic interactions, the string fragmentation model FRITIOF where we initialize a zero mass $\rho^0$-meson for the photon following a VMD picture. For the total cross section we use a parameterization from Ref.~\cite{PDG94}. The Lund model is then used to determine the probabilities for the different final states. In Fig.~\ref{chargemult} we show that this gives a very good description of charged particle multiplicities in photon-proton collisions; the agreement seen there is better than could be expected from
a model that had been developed for applications at high energies.
However, we do not expect the Lund model to give correct predictions for
all specific channels, especially with respect to isospin. 
The role of the Lund model for our calculations
is to supply us with an overall description of the elementary reaction 
dynamics in order to allow us to take into account multi-step processes where,
for example, a primary produced pion produces a vector meson on a second
nucleon.  
\par
\begin{figure}[h]
\centerline{\psfig{figure=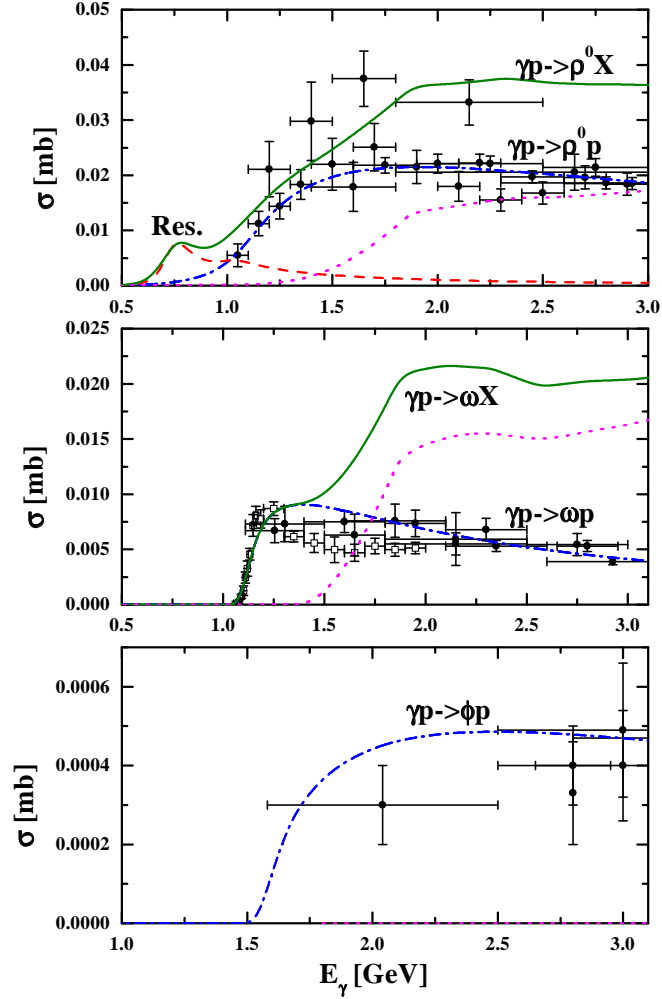,width=10cm}}
\caption{The $\rho$ (upper part), $\omega$ (middle part) and $\phi$
(lower part) meson production cross section for $\gamma p$ reactions.
The experimental data corresponding to the exclusive reactions $\gamma
p\to V p$ ($V=\rho,\omega,\phi$) are taken from
Ref.~\protect\cite{ABBHHM} (full circles) and from
Ref.~\protect\cite{expomegaN} (open squares). The dash-dotted lines are
our parametrization of the exclusive data; the dashed line (upper part)
is the resonance contribution. The dotted lines indicate the calculated
inclusive vector meson production cross section(see text); the
solid lines correspond to the total vector meson production cross
section.}
\label{Fig1gam}
\end{figure}

 The vector meson ($\rho,\omega,\phi$) production in $\gamma
N \to V N$ collisions is fitted to experimental data~\cite{ABBHHM,expomegaN} and treated
independent of the Lund model also for high energies. The cross section is given as:
\begin{equation}
\sigma_{\gamma N \to NV} = {1\over p_i s} \int d\mu |{\cal M}_{V}|^2 p_f
{\cal A}_V(\mu) \quad,
\label{gamNV}\end{equation}
where $\sqrt{s}$ is the total energy of the $\gamma N$ system, $p_i, p_f$
are the momenta of the initial and final particles in the center-of-mass
system, and ${\cal A}_V$ is the spectral function of vector meson $V$ (Eq.~\ref{spectral}).
The matrix elements ${\cal M}_V$ are parameterized in the following way:
\begin{eqnarray}
&&|{\cal M}_\rho|^2=0.16 \ \ {\rm mb\cdot GeV}^2 \label{matrel} \\
&&|{\cal M}_\omega|^2={0.08 p_f^2\over 2(\sqrt{s}-1.73{\rm GeV})^2 + p_f^2}
  \ \ {\rm mb\cdot GeV}^2 \nonumber \\
&&|{\cal M}_\phi|^2=0.004 \ \ {\rm mb\cdot GeV}^2 \quad.\nonumber
\end{eqnarray}
In Fig.~\ref{Fig1gam} we show the resulting cross sections (dash-dotted curves) for $\gamma p\to p\rho^0$ (upper part), $\gamma p\to p\omega$ (middle part) and $\gamma p\to p\phi$ (lower part) in comparison with the experimental data. For the angular distribution of the produced vector mesons we use
\begin{equation}
\frac{d\sigma}{dt} \propto \exp(Bt) \quad,
\label{angdist}
\end{equation}
where $t$ denotes the square of the 4-momentum transfer of the photon to the vector meson. In Ref.~\cite{ABBHHM} the parameter $B$ was, dependent on photon energy, fitted to $\rho^0$-production. Here, we adopt these values and also use them for $\omega$- and $\phi$-production.    
\par In our calculations there is an additional contribution to $\gamma N \to N \rho^0$ coming from the decays of the $N(1520)$ and $N(1680)$ resonances which is shown in Fig.~\ref{Fig1gam} by the dashed line. These decays predominantly contribute to low mass $\rho$-mesons below the experimentally seen $\rho$ production threshold.  
\par Besides the exclusive process $\gamma N \to V N$ we also have, for the photon energies considered here, to take into account additional channels for photoproduction of vector mesons. 
For energies above 2.1 GeV we calculate these cross sections using the Lund
model. Below 2.1 GeV we absorb everything into the channel $\gamma N \to V \pi N$ for which we use the following cross section:
\begin{equation}
\sigma_{\gamma N \to V \pi N}=\frac{16(2 \pi)^7}{p_i \sqrt{s}} \int d\mu d\Phi_3 |{\cal M}_{V\pi}|^2 {\cal A}_V(\mu) \quad,
\label{gamNV2}
\end{equation}
where $d\Phi_3$ denotes the 3-body phase space element as, for example, given by Eq. (35.11) in Ref.~\cite{PDG2}. The matrix elements are adjusted to give a continous transition to the string fragmentation model at $\sqrt{s}=2.1\,{\rm GeV}$. We use:
\[
|{\cal M}_{\rho^0 \pi}|^2=|{\cal M}_{\omega \pi}|^2=0.5\,{\rm mb} \quad.
\]  
The calculated inclusive rho and omega production cross sections (after
the subtraction of the exclusive part) are shown as dotted lines in
Fig.~\ref{Fig1gam}. The total inclusive vector meson production cross
sections are indicated as solid lines.
\par For photon energies above 1 GeV we take into account nuclear shadowing
of the incoming photon by adopting the model of Ref.~\cite{boffi}. Since 
this has, for the photon energies considered here, only a small impact on
the final results our specific transport theoretical realization will be
described elsewhere \cite{prodpaper2}. 

\subsection{Dilepton production}

In our analysis we calculate dilepton production by taking into account
the contributions from the Dalitz-decays $\Delta \to N e^+e^-$, $\eta
\to \gamma e^+e^-$, $\omega \to \pi^0 e^+e^-$, $\pi^0 \to \gamma e^+ e^-$ and the direct dilepton decays of the vector mesons $\rho, \omega, \phi$. 
\par The Dalitz decays of the $\pi^0$ and the $\eta$ are parameterized according to Ref.~\cite{landsberg}. For the Dalitz decay of the $\omega$ we use the parameterization from Ref.~\cite{Brat97}.
The $\Delta$ Dalitz-decay is described in line with Ref.~\cite{Wolf90} where we, however, use $g = 5.44$ for the coupling constant in order to reproduce the photonic decay width $\Gamma_0 (0) = 0.72$~MeV.
\par The dilepton decay of the vector mesons is calculated assuming strict vector meson dominance \cite{Sakurai69} as in Ref.~\cite{likobrown}:
\begin{equation}
\Gamma_{V \to e^+e^-}(M) = C_V {m_V^4\over M^3},
\label{gee}\end{equation}
where $C_\rho = 8.814\times 10^{-6}$, $C_\omega = 0.767\times 10^{-6}$
and $C_\phi = 1.344\times 10^{-6}$, respectively \cite{PDG2}. Within an extended vector meson dominance picture \cite{evmd} one 
has a dilepton decay amplitude that consists of 2 terms, one describing the
coupling of the virtual photon to the hadron with a strength proportional to
$M^2$ and another with strength proportional to $M^0$.
However, since we neglect a direct coupling of the virtual photon the use of strict vector meson dominance is more appropriate. We also note that this dilepton decay width together with our parameters for the $\rho$-meson gives a very good description of the experimental data for $e^+ e^- \to \pi^+ \pi^-$.  
\par In our transport model the dilepton yield is obtained from the 
phase space distributions of the respective sources by a time integration.
For the vector mesons the massdifferential dilepton production is given
as:
\begin{equation}
\frac{dN_{V \to e^+ e^-}}{d \mu} = \int_0^\infty dt \ d^3r \ 
\frac{d^3 p}{(2 \pi)^3} \ F_V(\vec{r},t,\vec{p},\mu) \ 
\frac{\Gamma_{V \to e^+ e^-}}{\gamma} \quad,
\end{equation}
where $\gamma$ is a Lorentz factor which appears since 
$\Gamma_{V \to e^+ e^-}$ is the width in the rest frame of the vector meson.   
The Dalitz decay contributions contain an additional mass integration. For the
$\Delta$-resonance we have for example:
\begin{equation}
\frac{dN_{\Delta \to N e^+ e^-}}{d \mu} = \int_0^\infty dt \ d^3r \ 
\frac{d^3 p}{(2 \pi)^3} \ d\mu_2 \ F_\Delta(\vec{r},t,\vec{p},\mu_2) \ 
\frac{1}{\gamma} \ \frac{d\Gamma_{\Delta \to N e^+ e^-}}{d \mu} .
\end{equation}   
  
\section{Dilepton production in $\gamma A$ reactions}

\begin{figure}[h]
\centerline{\psfig{figure=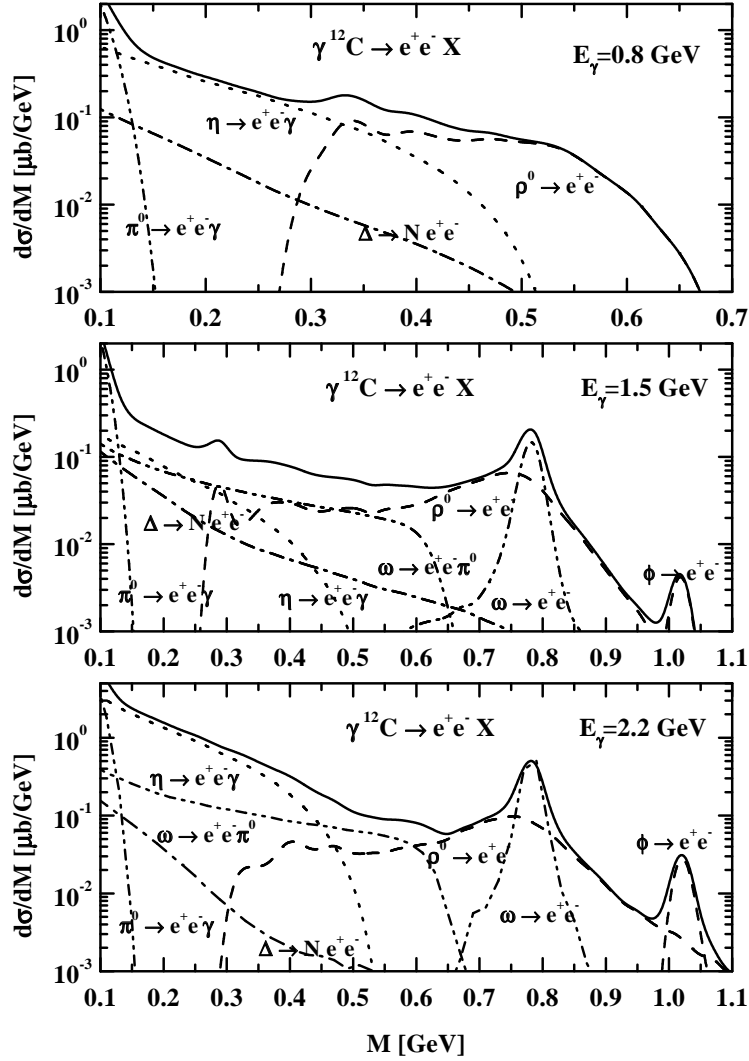,width=12cm}}
\caption{
The dilepton invariant mass spectra $d\sigma/dM$ for $\gamma$C at the
energy of $E_\gamma =0.8$~GeV (upper part), 1.5~GeV (middle part)
and 2.2~GeV (lower part) calculated with bare meson masses
including a mass resolution of 10~MeV.
The thin lines indicate the individual contributions from the different
production channels; {\it i.e.} starting from low $M$: Dalitz decay
$\pi^0 \to \gamma e^+ e^-$ (short-doted line),
$\eta \to \gamma e^+ e^-$ (dotted line), $\Delta \to N e^+ e^-$
(dot-dashed line), $\omega \to \pi^0 e^+ e^-$ (dot-dot-dashed line);
for $M \approx $ 0.8 GeV:  $\omega \to e^+e^-$ (dot-dot-dashed line),
$\rho^0 \to e^+e^-$ (dashed line).  The full solid line represents
the sum of all sources. }
\label{Fig2gam}
\end{figure}

\begin{figure}[h]
\centerline{\psfig{figure=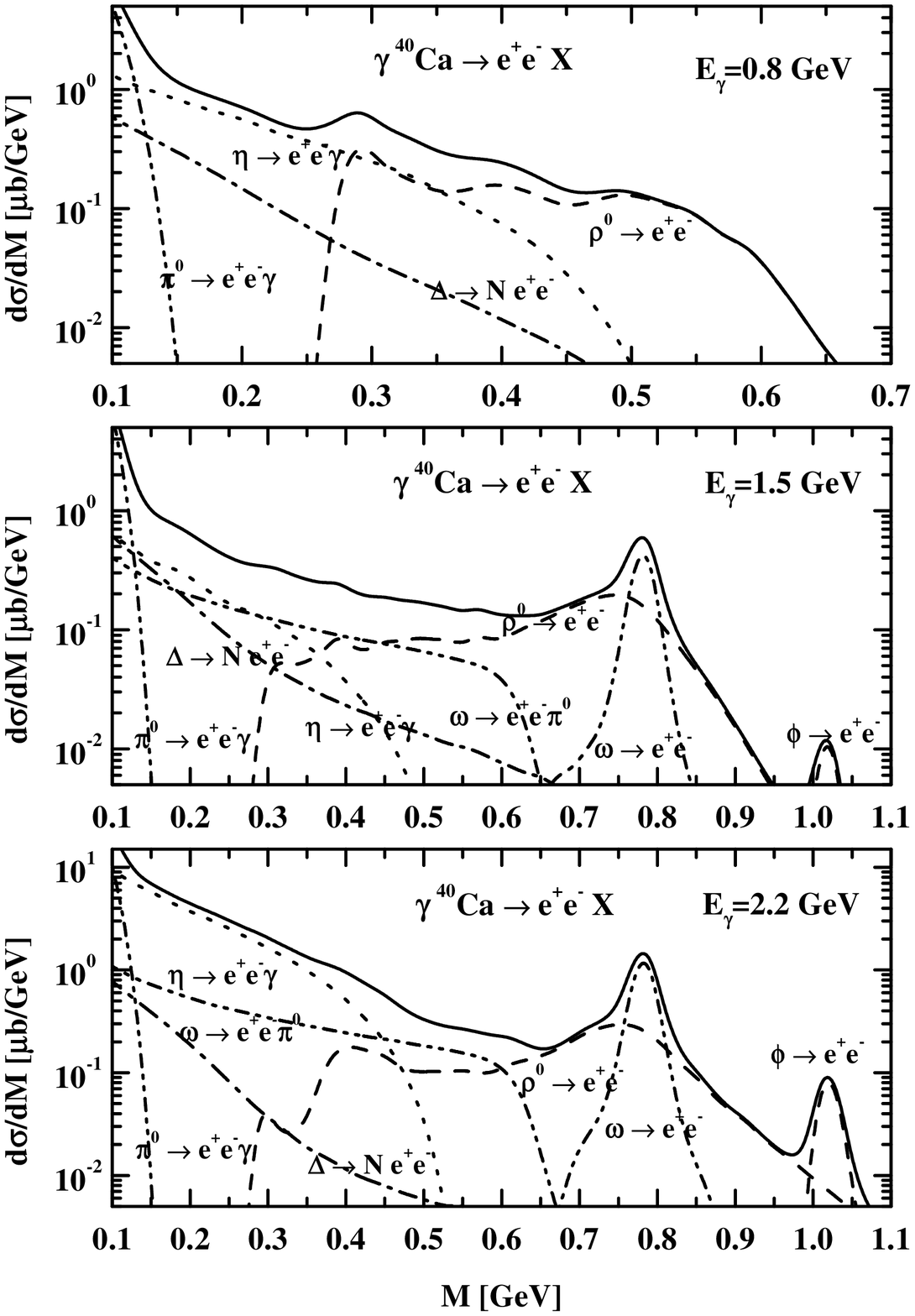,width=12cm}}
\caption{ The dilepton invariant mass spectra $d\sigma/dM$ for
$\gamma$Ca at the energy of $E_\gamma =0.8$~GeV (upper part),
1.5~GeV (middle part) and 2.2~GeV (lower part) calculated with bare meson
masses including a mass resolution of 10~MeV. The assignment is the
same as in Fig.~\ref{Fig2gam}.}
\label{Fig3gam}
\end{figure}

\begin{figure}[h]
\centerline{\psfig{figure=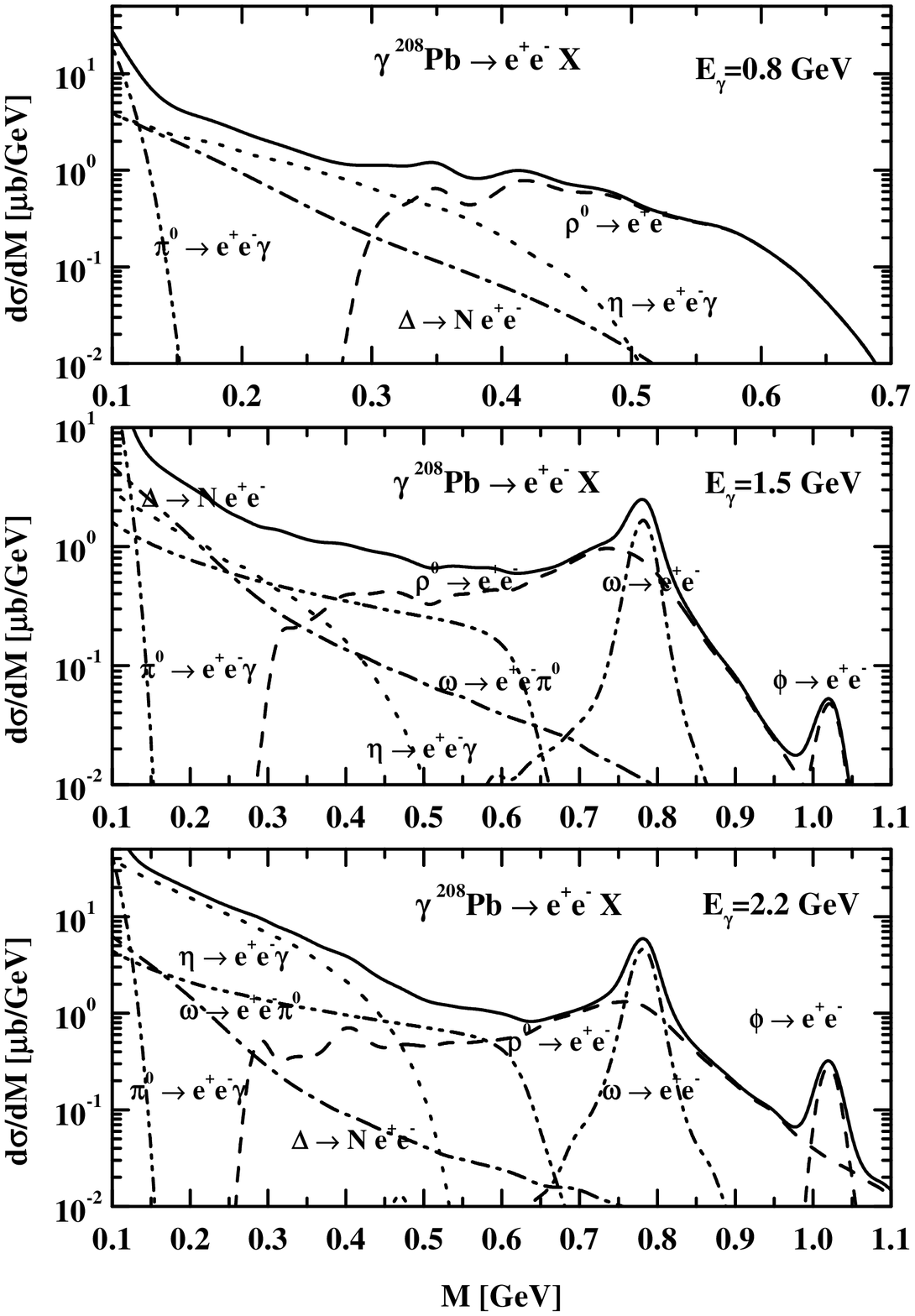,width=12cm}}
\caption{ The dilepton invariant mass spectra $d\sigma/dM$ for
$\gamma$Pb at the energy of $E_\gamma =0.8$~GeV (upper part),
1.5~GeV (middle part) and 2.2~GeV (lower part) calculated with bare meson
masses including a mass resolution of 10~MeV. The assignment is the
same as in Fig.~\ref{Fig2gam}.}
\label{Fig4gam}
\end{figure}

In Figs.~\ref{Fig2gam}--\ref{Fig4gam} we present the calculated
dilepton spectra $d\sigma/dM$ for $\gamma$C, $\gamma$Ca,
$\gamma$Pb reactions at photon energies $E_\gamma = 0.8, \ 1.5, \
2.2$~GeV. A mass resolution of 10~MeV is included through a convolution of our calculated spectrum with a Gaussian. Here neither collisional broadening nor a mass shift of the vector mesons were taken into account.
\par The thin lines indicate the individual contributions from the different
production channels; {\it i.e.} starting from low $M$: Dalitz decay
$\pi^0 \to \gamma e^+ e^-$ (short-doted line),
$\eta \to \gamma e^+ e^-$ (dotted line), $\Delta \to N e^+ e^-$
(dot-dashed line), $\omega \to \pi^0 e^+ e^-$ (dot-dot-dashed line);
for $M \approx $ 0.8 GeV:  $\rho^0 \to e^+e^-$ (dashed line),
$\omega \to e^+e^-$ (dot-dot-dashed line), $\phi \to e^+e^-$ (dashed line). 
The full solid lines represent the sum of all sources.
The dominant processes in the low mass region up to
$M\simeq 500$~MeV are the $\eta$, $\omega$ and $\Delta$ Dalitz decays.
Above $M\sim 0.6$~GeV the spectrum is dominated by the vector meson
decays with a low background from other hadronic sources.
\par In our calculations we only take into account $\rho$-mesons with masses
above $2 m_\pi$ which is the threshold of the strong decay since in the
calculation of the spectral function (Eq.~(\ref{spectral})) we
neglect contributions from electroweak decays. Therefore we get a
discontinuity of our spectra in Figs.~\ref{Fig2gam}--\ref{Fig4gam} at the
two-pion mass that is, however, because of the other sources and the mass
resolution hardly visible.  

\subsection{Bethe-Heitler contribution}

\begin{figure}[h]
\centerline{\psfig{figure=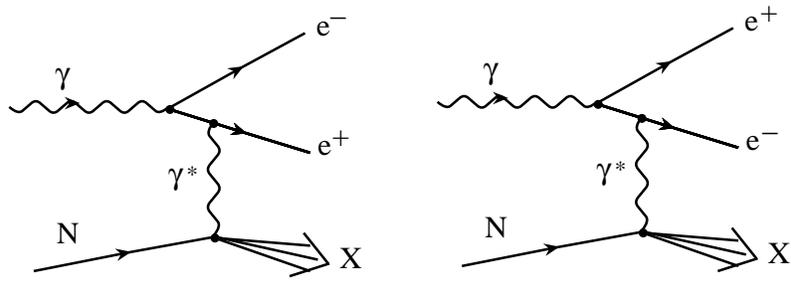,width=12cm}}
\caption{Feynman diagram for $\gamma N \to e^+e^- X$ for the
Bethe-Heitler process.}
\label{Fig5gam}
\end{figure}

Besides the 'hadronic' contributions as discussed above we also have to take into account dilepton production via the so-called Bethe-Heitler (BH) process for which the Feynman diagrams are depicted in Fig.~\ref{Fig5gam} that contribute to lowest order in the electromagnetic coupling constant $\alpha$. 

On a single nucleon, with the electromagnetic form factors known from electron scattering, the BH process is completely determined by QED dynamics and can easily be calculated. For a detailed description of the involved matrix elements we refer to Ref.~\cite{Tsai} from which we also adopted the parameterizations of the electromagnetic form factors $W_1(Q^2,\nu)$ and $W_2(Q^2,\nu)$ of the nucleon.
\par For our calculations we take only the incoherent sum over BH contributions on single nucleons into account and neglect contributions where the intermediate photon couples to the charge of the whole nucleus. While the latter will, because of the $Z^2$ dependence, dominate all integrated cross sections it can experimentally easily be suppressed by appropriate missing mass cuts.
\par
\begin{figure}[h]
\centerline{\psfig{figure=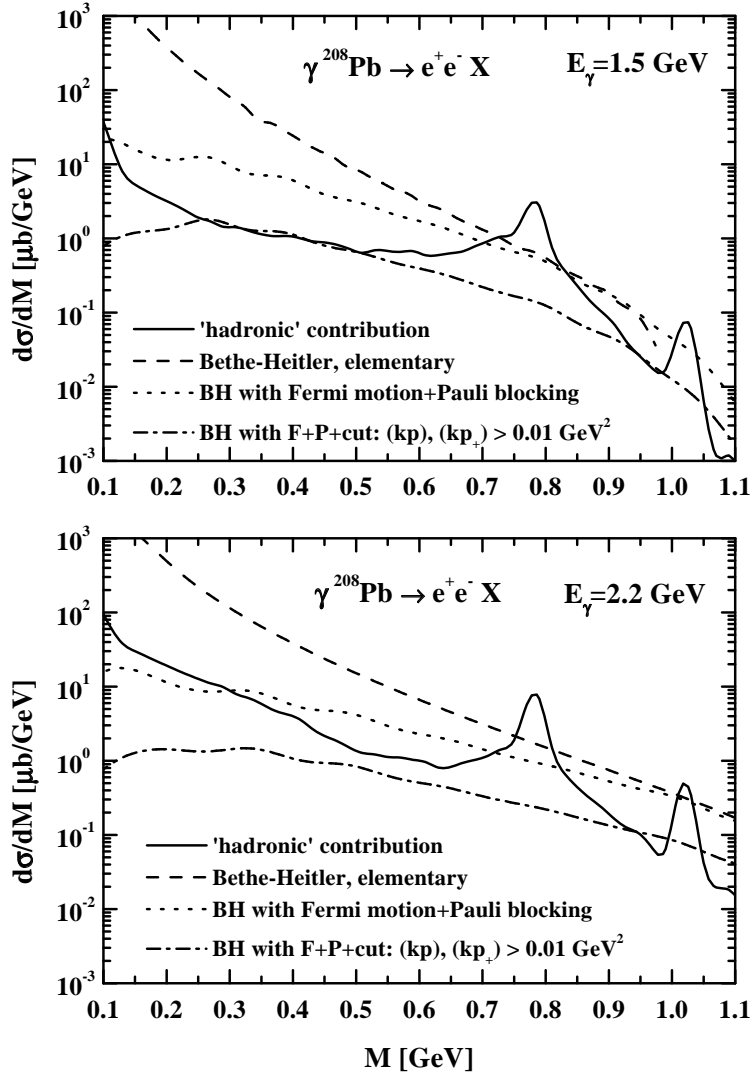,width=12cm}}
\caption{The dilepton invariant mass spectra $d\sigma/dM$ for
$\gamma$Pb at the energy of $E_\gamma =1.5$~GeV (upper part)
and 2.2~GeV (lower part). The solid lines indicate the 'hadronic'
spectra as in Fig.~\ref{Fig4gam}. The dashed lines are the
Bethe-Heitler contribution calculated as a sum of incoherent
contributions from $\gamma$+nucleons.  The dotted lines show the BH yield
with taking into account nucleon Fermi motion and Pauli blocking.
The dot-dashed lines are the BH terms with the cuts
$(k \cdot p), (k \cdot p_+) > 0.01$~GeV. }
\label{Fig6gam}
\end{figure}

 In Fig.~\ref{Fig6gam} we compare the BH contributions for $\gamma$Pb at 1.5 (upper part) and 2.2 GeV (lower part) with the 'hadronic' contributions that we have already shown in Fig.~\ref{Fig4gam}. One sees that the sum of the elementary cross sections on the proton and the neutron (dashed lines) is much larger than the 'hadronic' contributions (solid lines) for dilepton masses below 0.6 GeV. In the region of the $\rho$ and $\omega$ meson the BH contribution is about a factor of 4 smaller but still non-negligible. With the inclusion of Fermi motion and Pauli blocking (dotted lines) the BH contribution is reduced significantly for low invariant masses but hardly affected for masses larger than 700 MeV. 
\par In order to suppress the BH contribution we
implemented the cuts $(k \cdot p), (k \cdot p_+) > 0.01$~GeV$^2$, where $k$ is
the 4-momentum of the incoming photon, and $p$, $p_+$ are the 4-momenta
of electron and positron, respectively. These cuts, which reflect the
pole-like behaviour of the intermediate electron propagator, suppress the BH contribution
by a factor of 10 -- dot-dashed lines in Fig.~\ref{Fig6gam} -- and
practically do not have any influence on the 'hadronic' contributions. 
\par In our calculations we do not take into account interference terms 
between the BH and the 'hadronic' contributions. 
For an inclusive cross section that contains to each $e^+ e^-$ pair the pair
with exchanged momenta the interference term vanishes and the BH 
contribution can easily be subtracted. Therefore we will in the following
discuss only the hadronic contributions. 

\section{In-medium effects in dilepton production}

\subsection{Collisional broadening}
\label{collbroad}
The in-medium widths of the $\rho$ and $\omega$ mesons are calculated as 
sketched in Section~\ref{broadres}. In the rest frame of the meson the total
in-medium width is given as:  
\begin{equation}
\Gamma^V_{tot}(\mu,|\vec{p}|,\rho)=\Gamma^V_{vac}(\mu)
+ \Gamma^V_{coll}(\mu,|\vec{p}|,\rho) \quad,
\label{gammas}
\end{equation}
where the collisional width $\Gamma^V_{coll}$ reads:
\begin{equation}
\Gamma^V_{coll}(\mu,|\vec{p}|,\rho) = \gamma \rho <v_{VN} \sigma_{VN}^{tot}> \quad,
\label{dgamma}
\end{equation}
and $\Gamma^V_{vac}$ is the vacuum decay width.
In Eq.~(\ref{dgamma}) the brackets stand for an average over the Fermi sea of the nucleons,
$v_{VN}$ is the relative velocity between vector meson and nucleon, and 
$\sigma_{VN}^{tot}$ is their total cross section. $\rho$ is the nuclear density
and $\gamma$ the Lorentz factor for the boost to the rest frame of the vector
meson.
\par
\begin{figure}[h]
\centerline{\psfig{figure=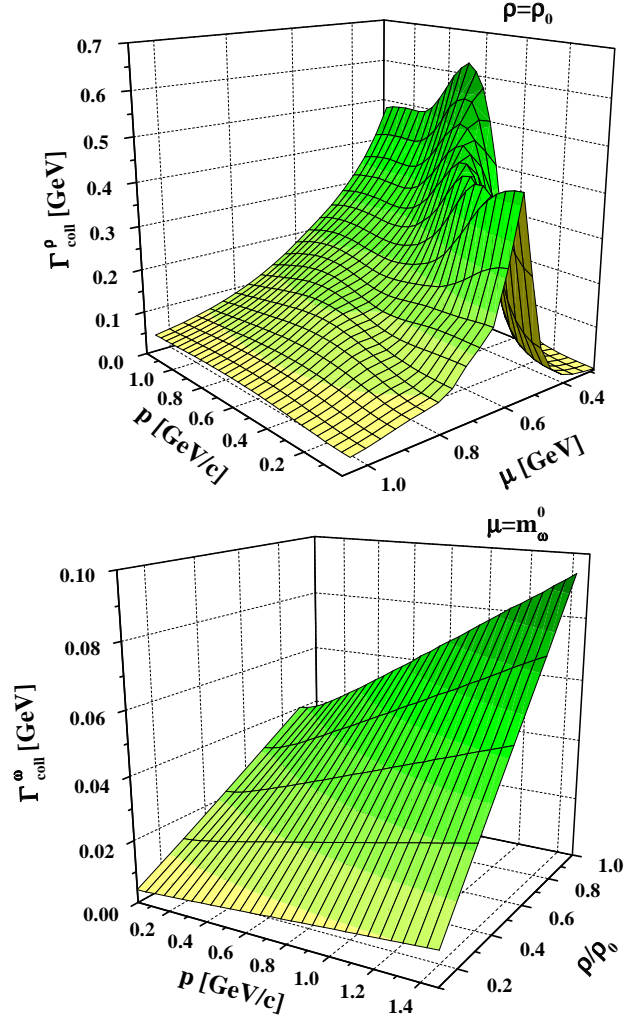,width=12cm}}
\caption{The upper part shows the collisional width of the $\rho$ meson
as a function of momentum and mass at normal nuclear density $\rho=\rho_0$.
The lower part is the momentum and density dependence of the $\omega$
collisional width calculated at $\mu=m_\omega^0$. }
\label{Fig7gam}
\end{figure}

 The upper part of
Fig.~\ref{Fig7gam} shows the collisional width of the $\rho$ meson as a
function of momentum and mass at nuclear matter density $\rho=\rho_0$.
The structure at low $\mu$ comes from the resonance contributions,
especially from the $D_{13}(1520)$. Note that the width becomes very large
(up to about 600 MeV), corresponding to a complete melting of the 
$\rho$-meson. 
The lower part of Fig.~\ref{Fig7gam}
shows the momentum and density dependence of the $\omega$
collisional width calculated at the pole mass $\mu=m_\omega^0$. At nuclear
matter density $\rho_0$ and a momentum of $p=1\,{\rm GeV}$ we obtain a collisional width of about 80 MeV which is about a factor of 10 larger than the
vacuum decay width.
\par
\begin{figure}[h]
\centerline{\psfig{figure=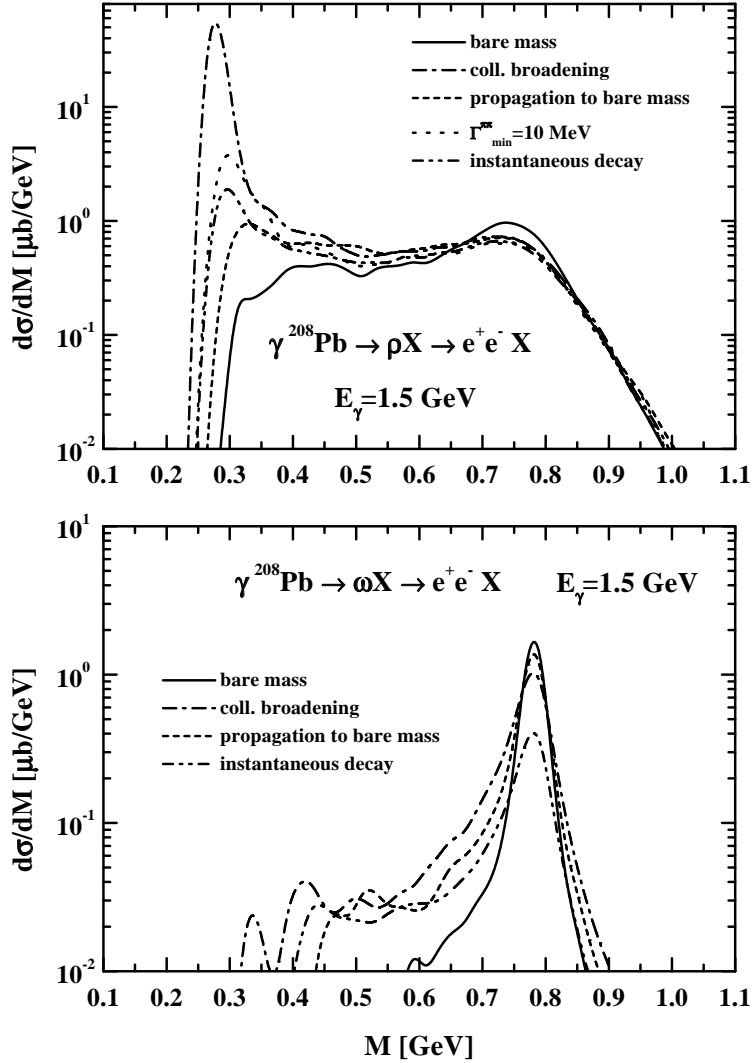,width=12cm}}
\caption{The dilepton yield from $\rho$ (upper part) and $\omega$
(lower part) meson decays for $\gamma$Pb at 1.5~GeV calculated within
different prescriptions: the solid lines are the results with bare mass,
the dot-dashed lines indicate the calculation with collisional broadening,
the dot-dot-dashed lines correspond to the instantaneous decay,
the dotted line (upper part) is the result with
$\Gamma_{min}^{\pi\pi}=10$~MeV, the short dashed lines indicate
the yield calculated with the 'propagation to bare mass' prescription.}
\label{Fig8gam}
\end{figure}

 In Fig.~\ref{Fig8gam} (upper part) we show the contribution of the $\rho$-meson to the
$e^+e^-$-yield for $\gamma$Pb at 1.5 GeV. The solid line indicates the bare
mass case (as in Fig.~\ref{Fig4gam}), i.e. without collisional broadening. The
curve labelled 'coll. broadening' (dash-dotted line) results if we take into account the collision broadening
effect in the production of the $\rho$-mesons. Here we calculate the 
$\rho$-meson production cross sections in photon-nucleon collisions
(Eqs.~(\ref{gamNV}),(\ref{gamNV2})) with the in-medium spectral function. 
Since the in-medium spectral function depends on the momentum of the
$\rho$-meson with respect to the nuclear medium an additional angular
integration has to be performed. For the case of exclusive production
$\gamma N \to N \rho$ we use the angular distribution from Eq.~(\ref{angdist}).
For $\gamma N \to N \rho \pi$ we assume an isotropic three-body phase space
distribution.  
\par As discussed in Section~\ref{broadres} we do not modifiy the $N\rho$-width
of the baryonic resonances but the masses of the $\rho$-mesons stemming from
these decays are distributed according to the (phase space weighted) in-medium
spectral function. We, again, neglect $\rho$-mesons with masses below the
two-pion threshold. 
\par From Fig.~\ref{Fig8gam} one sees that the inclusion of
collisional broadening leads to a depletion of the $\rho$-meson peak by about
30\% and a shift of strength to lower dilepton masses. One also observes a
very large peak at $M=2m_\pi$ which is in fact a pole but here finite due to
our numerical solution. The reason for this divergence is directly 
related to our discussion in Section~\ref{broadres} of the asymptotic 
solutions of the semi-classical transport equation when including 
in-medium spectral functions. At the two-pion mass the vacuum spectral function of the $\rho$-meson is zero while the in-medium spectral
function has some finite value since the collision width from 
Eq.~(\ref{dgamma}) does not vanish. When travelling to the vacuum the respective component of the spectral phase space density becomes infinitely long lived
and leads to the divergence. If we included the electroweak decay width
of the $\rho$-meson into the collision term of the transport equation the
problem would not be solved. The pole would only be replaced by a numerically
indistinguishable large peak.
\par In Fig.~\ref{Fig8gam} (lower part) we show the contribution of the
$\omega$-meson when including collisional broadening (dash-dotted line) in
comparison to the calculation with the vacuum spectral function (solid line, as in Fig.~\ref{Fig4gam}). One observes a strong broadening of the 
$\omega$-peak which is partly covered up by the inclusion of a mass
resolution of 10 MeV. However, such a strong broadening is not realistic
because most of the $\omega$-mesons that contribute to the dilepton spectrum
decay in the vacuum. This is also 
reflected by the violation of the condition Eq.~(\ref{condition}) for the
validity of our off-shell transport equation in this case.

\subsubsection{Prescriptions to obtain correct asymptotic behaviour}
\label{recipes}
In the following we want to discuss a prescription that 
allow us to obtain a divergence free $\rho$-meson and a reasonable 
$\omega$-meson contribution. For that purpose we introduce a potential that
shifts a particle to its vacuum spectral function when it propagates to the
vacuum. Such a potential can not be defined on the level of the transport
equation Eq.~(\ref{transport}). However, we can introduce such a potential
on the level of our numerical realization. We recall that we solve the
transport equation by a so-called test particle method, i.e. we make an ansatz
for the spectral phase space density $F$:
\begin{equation}
F(\vec{r},t,\vec{p},\mu) \propto \sum_i \delta(\vec{r}-\vec{r}_i(t))
\delta(\vec{p}-\vec{p}_i(t)) \delta (\mu-\mu_i)  \quad,
\label{testpar}
\end{equation}
where $\vec{r}_i(t)$, $\vec{p}_i(t)$, and $\mu_i$ denote the spatial
coordinate, the momentum and the mass of the test particle $i$, respectively.
Now, we can define for each test particle a density dependent scalar potential
$s_i$ in the following way:
\begin{equation}
s_i(\rho_i(t))=(\mu^{med}_i-\mu^{vac}_i)\frac{\rho_i(t)}{\rho^{cr}_i} \quad,
\label{scalP}
\end{equation} 
where $\mu^{med}_i$ is the 'in-medium' mass of the test particle chosen 
according to the mass differential production cross section. $\mu^{vac}_i$
is the 'vacuum' test particle mass which is chosen according to the production
cross section with a vacuum spectral function. 
$\rho^{cr}_i$ is the baryon density at the creation point, whereas
$\rho_i(t)=\rho(\vec{r}_i(t))$ is the baryon density during the propagation.
Setting the test particle mass $\mu_i$ to the 'vacuum' mass $\mu^{vac}_i$ the
effective in-medium mass $\mu^\star_i$ is given as:
\begin{equation}
\mu^\star_i(\rho_i(t)) = \mu^{vac}_i + s_i(\rho(t)) \quad.
\label{scalMass}
\end{equation}
Eq.~(\ref{scalMass}) gives the correct asymptotic behaviour:  the
effective mass at the creation point is kept unchanged
($\mu^\star_i(\rho^{cr}_i)=\mu^{med}_i$); 
during the propagation the test particle mass
changes linearly with density and outside the nucleus it becomes
equal to the bare mass $\mu^\star_i(\rho_i=0)=\mu^{vac}_i$.
The potential $s_i$ enters into the test particle propagation as a
usual potential and therefore guarantees that this prescription does not violate energy conservation.
\footnote{After this paper was submitted two other papers dealing with the
propagation of broad resonances appeared on the preprint server 
\cite{cassing,knoll2}.}
\par The potential Eq.~(\ref{scalP}) can assume rather large values in the
case of broad resonances even if the in-medium corrections are small. Its
effect is, nevertheless, in this case negligible. As a check
we have performed a calculation of photoproduction of 
$\rho$-mesons in which we distributed the masses $\mu^{med}_i$ according to 
the vacuum spectral function. This gave practically the same result as the 
calculation without potential since the lifetime of the $\rho$-mesons is so
short that only very few propagate through a relevant density gradient for 
which the potential becomes important.
\par In Fig.~\ref{Fig8gam} we show the results of our calculations with
the described prescription for the $\rho$ and the $\omega$ mesons (curves 
labelled 'propagation to bare mass', dashed lines). For the $\rho$-meson
the divergence at the two-pion threshold is removed. For invariant masses
above 500 MeV we get practically the same result as without the prescription because
of the reason described in the preceding paragraph. The broadening of the
$\omega$-peak is reduced because now only the $\omega$-mesons that decay
inside the nucleus contribute to the broadening.
\par Another possibility to overcome the problems with the transport theoretical treatment of broad resonances is to avoid their explicit propagation.
Therefore we have also performed simulations in which we calculated the cross sections
for elementary dilepton production via vector mesons as instantaneous one-step
processes. The results are shown in Fig.~\ref{Fig8gam} by the curves
labelled 'instantaneous decay' (dot-dot-dashed lines). 
The $\rho$-meson contribution is almost identical to the one calculated with
the prescription described above because of its very short life time. 
For the $\omega$-meson we get a reduction by about a factor of 3 since 
in the instantaneous decay scheme
we neglect the possibility that an $\omega$-meson can escape from the nucleus
and then contribute to dilepton production with a much larger branching ratio
than inside the nucleus. As such dynamical effects are important to take
into account we consider the description via an instantaneous process to be
inadequate.              
\par The easiest way to cope with the divergence of the $\rho$-meson 
contribution is to use a minimum two-pion decay width. In Fig.~\ref{Fig8gam}
we show the result of a calculation where we set this minimum width to
10 MeV (dotted line). One sees that still a large peak remains. Moreover, this 
approach is in any case questionable.
\par For the reasons described above we consider the prescription 'propagation to bare mass' as the only possibilty to implement collision broadening effects in
our calculations. However, we want to stress that this prescription is
not fully satisfactory since it is only
formulated on the level of our specific numerical solution of the transport
equation and since it neglects any memory effects.

\subsection{'Dropping' vector meson mass}

In order to explore the observable consequences of vector meson mass
shifts at finite nuclear density the in-medium vector meson masses are
modeled according to Brown/Rho scaling
\cite{BrownRho} or Hatsuda and Lee \cite{H&L92} by introducing a scalar potential $S_V(\vec{r})$:
\begin{equation}
\label{Brown}
S_V(\vec{r})=-\alpha m_V^0 \frac{\rho(\vec{r})}{\rho_0} \quad,
\end{equation}
where $\rho(\vec{r})$ is the nuclear density, $m_V^0$ the pole mass of the vector meson, $\rho_0 = 0.168 \ {\rm fm}^{-3}$, and $\alpha \simeq 0.18$ for the $\rho$ and $\omega$. The effective mass $\mu^\star$ is then given as:
\begin{equation}
\mu^\star=\mu+S_V \quad.
\end{equation}
For the effective pole mass $\mu^\star_0$ we thus get:
\[\mu^\star_0=\left( 1-\alpha \frac{\rho(\vec{r})}{\rho_0} \right) m_V^0 \quad. \] 
\par In photon-nucleus reactions vector mesons are produced with large momenta relative to the nuclear medium. Within a resonance-hole model for the $\rho$-meson self energy in the nuclear medium it has been shown in Ref.~\cite{Kond98rho} that the real part of the in-medium $\rho$-meson self energy increases with momentum and crosses zero for a momentum of about 1 GeV. In order to explore the implications of such a behaviour we also use a momentum dependent scalar potential $S_V^{mom}$ for the vector mesons:
\begin{equation}
S_V^{mom}(\vec{r},\vec{p})=S_V(\vec{r})\left(1-\frac{|\vec{p}|}{1\, {\rm GeV}} \right) \quad.
\label{mompot}
\end{equation}
In our calculations we take the vector meson potentials into account for the calculation of the phase space factors in $\gamma N \to N V$ (Eq.~(\ref{gamNV})) and $\gamma N \to N V \pi$ (Eq.~(\ref{gamNV2})). We neglect these modifications for the vector meson production in the string fragmentation model FRITIOF and also do not modify the $N \rho$-widths of the baryonic resonances. 
 
\subsection{Dileptons from $\gamma A$ reactions: In-medium effects}

\begin{figure}[h]
\centerline{\psfig{figure=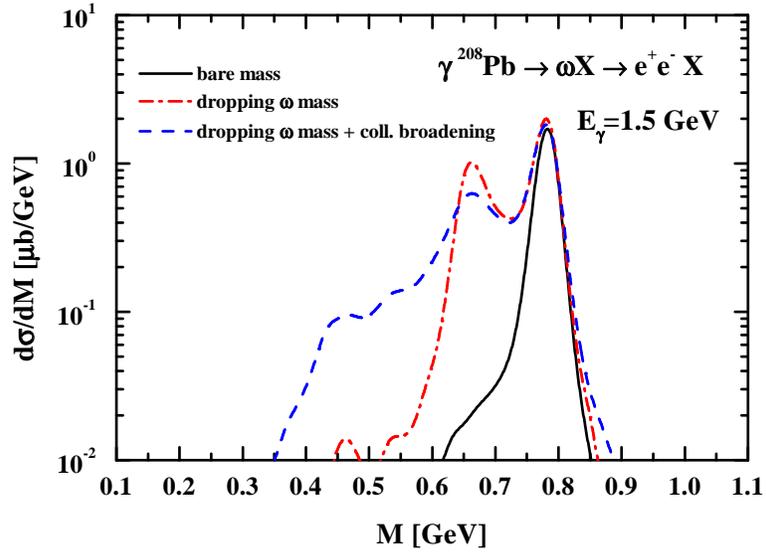,width=12cm}}
\caption{The dilepton yield from $\omega$ mesons for $\gamma$Pb at 1.5 GeV.
The solid line indicates the bare mass case, the dot-dashed line is the
result with in-medium masses, the dashed line shows the effect of
collisional broadening together with the dropping mass.}
\label{Fig9gam}
\end{figure}

In Fig.~\ref{Fig9gam} we show the contribution coming from the $\omega$-meson
to $e^+e^-$-production in $\gamma$Pb at a photon energy of 1.5 GeV.
A dropping mass scenario according to Eq.~(\ref{Brown}) (dot-dashed line)
gives a two peak structure corresponding to $\omega$-mesons decaying
inside and outside the nucleus, respectively. An additional inclusion of
collisional broadening, as described in Section \ref{collbroad} gives a
substantial broadening of the dilepton yield from $\omega$-mesons that decay
inside the nucleus. The height of the peak around the vacuum pole mass of the
$\omega$-meson is hardly affected by the dropping mass scenario. On the one
hand the lowering of the mass reduces the vacuum peak because 
$\omega$-mesons decaying inside the nucleus contribute to lower masses. On the
other hand the total production of $\omega$-mesons is enhanced since the 
phase space factors entering the elementary photoproduction cross sections
(Eqs.~(\ref{gamNV}), (\ref{gamNV2})) are increased for lower masses. In our
calculation both effects nearly cancel each other for masses around the vacuum pole.
\par
\begin{figure}[h]
\centerline{\psfig{figure=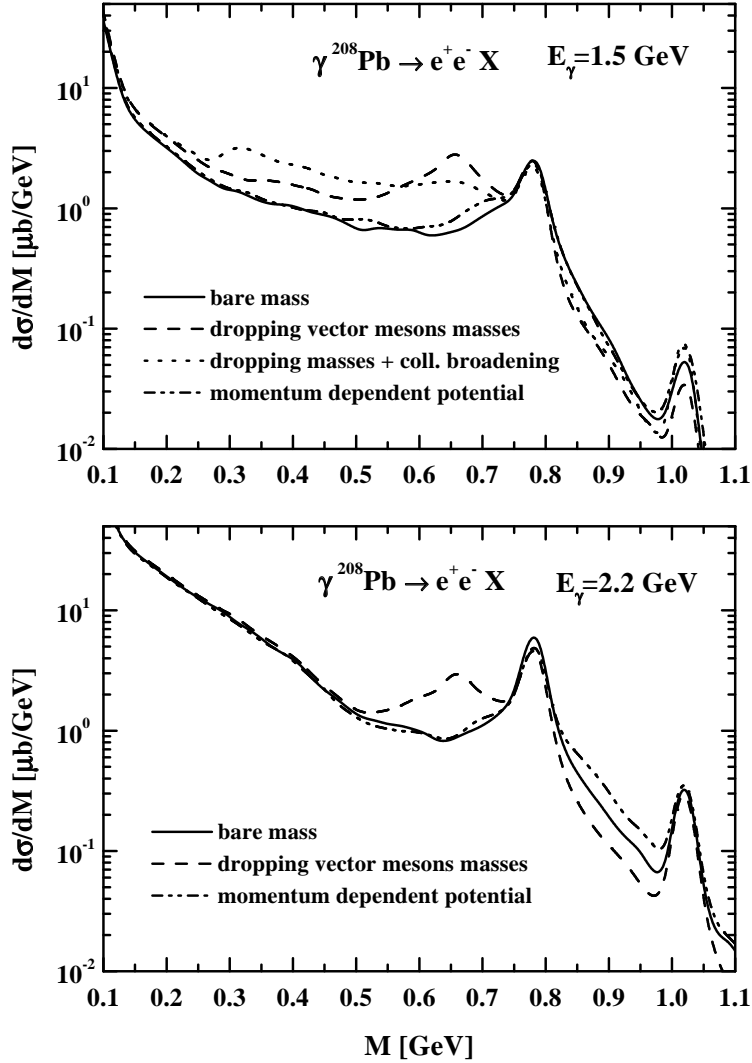,width=12cm}}
\caption{The dilepton invariant mass spectra for $\gamma$Pb at 1.5 GeV
(upper part) and 2.2 GeV (lower part).
The solid lines indicate the bare mass case, the dashed lines are the
result with the dropping mass scenario, the dotted line (upper part)
shows the effect of collisional broadening together with the dropping mass,
the dot-dot-dashed lines indicate the result with the momentum dependent
potential from Eq.~(\ref{mompot}).}
\label{Fig10gam}
\end{figure}

 In Fig.~\ref{Fig10gam} (upper part) we show the total $e^+e^-$ yield
for the same reaction. A dropping mass scenario for the vector mesons
(dashed line) leads to a second peak structure at invariant masse of about
650~MeV. The peak around 780~MeV remains practically unchanged since it is
dominated by $\omega$-mesons decaying outside the nucleus.
\par With the inclusion of collisional broadening the in-medium peak gets
completely washed out (dotted line). The dilepton yield at intermediate
masses is about a factor of 2 larger than in the bare mass case. At the
two-pion threshold there is a visible discontinuity which results from our
neglect of $\rho$-mesons with effective masses below the two-pion mass.
\par Using the momentum dependent scalar potential from Eq.~(\ref{mompot}) we
obtain the curves labelled 'momentum dependent potential' 
(dot-dot-dashed lines). 
The result is very close to the bare mass case as the vector mesons are
mainly produced with momenta around 1~GeV for which the potential is zero.
\par In the lower part of Fig.~\ref{Fig10gam} we show that the effect of a
dropping vector meson mass at a photon energy of 2.2 GeV is qualitatively the
same as for 1.5 GeV. The calculation with a momentum dependent potential
gives again a result that is practically the same as for the bare mass case.
\par In photonuclear reactions 
vector mesons are in general
produced with larger momenta relative to the nuclear medium than in heavy-ion
collisions. Since the in-medium spectral functions of the vector mesons are
momentum dependent one might thus observe rather different
in-medium effects in both reactions. These, together with
additional information from pion-nucleus and proton-nucleus reactions
\cite{HADES}, might help to discriminate between different
scenarios of medium modifications. Therefore a calculation of all
reactions within one model is necessary for a conclusive interpretation of 
the experimental data. Our BUU transport model provides such a tool.

\section{Summary}

We have studied $e^+e^-$ production in $\gamma$C, $\gamma$Ca, and $\gamma$Pb  
reactions at photon energies of 0.8, 1.5, and 2.2 GeV
within a semi-classical transport model. Various contributions were taken
into account for dilepton production: Dalitz decays of $\Delta(1232)$, $\pi^0$,
$\eta$, and $\omega$ as well as direct dilepton decays of the vectors mesons
$\rho$, $\omega$, and $\phi$. 
We have focused on observable effects of in-medium modifications of the
vector mesons $\rho$ and $\omega$. 
\par It was shown that the Bethe-Heitler process which dominates all
integrated cross sections for dilepton production can be sufficiently
suppressed by appropriate cuts on the lepton momenta. For dilepton invariant
masses above 600~MeV the spectrum is dominated by the the decays of the
vector mesons ($\rho$, $\omega$, $\phi$). A mass shift of these
mesons as proposed in Refs.~\cite{BrownRho,H&L92} leads to a substantial 
enhancement of the dilepton yield at invariant masses of about 650 MeV by
about a factor of 3 and should clearly be visible in experiments that will
be carried out at TJNAF~\cite{CEBAF}.
However, a calculation
for which we used a linearly momentum dependent potential for the vector
mesons \cite{Kond98rho} gave practically no effect compared to the bare mass
case. 
\par We have stressed the necessity of a simultaneous description of vector
meson production in different nuclear reactions as one probes in-medium
properties at different momenta relative to the nuclear medium. 
\par Exemplarily for the case of the $\rho$ and the $\omega$ mesons we have
discussed the conceptual problems in the treatment of broad resonances in
semi-classical transport models. We have presented a prescription that allows to
obtain reasonable results when taking into account in-medium spectral 
functions. 
\acknowledgements
The authors are grateful for discussions with W. Cassing.
This work was supported by DFG and GSI Darmstadt.


\newpage
\begin{table}[h]
\begin{center}
\caption{\label{mestable} Properties of mesonic resonances. $M_0$ and $\Gamma_0$ denote the pole mass and the width at the pole mass, respectively.}
\vspace{0.6cm}
\begin{tabular}{|c|c|c|c|}
Meson&$M_0$[MeV]&$\Gamma_0$[MeV]&decay channels\\
\hline
$\rho$&770&151&$\pi \pi$\\
$\omega$&782&8.4& $\pi \pi$ (2\%), $\pi^0 \gamma$ (9\%), $\pi^+ \pi^- \pi^0$ (89\%)\\
$\phi$&1020&4.4&$\rho \pi$ (13\%), $K \bar{K}$ (84\%), $\pi^+ \pi^- \pi^0$ (3\%)\\
$\sigma$&800&800&$\pi \pi$\\
\end{tabular}
\end{center}
\end{table}

\newpage

\end{document}